# SEMIDV: A Compact Semiconductor Device Simulator with Quantum Effects

Chien-Ting Tung

*Abstract*—In this paper, I present SEMIDV – a compact semiconductor device simulator incorporating quantum effects. Localization landscape theory is introduced to provide quantum corrections to the Drift-Diffusion equation. This theory directly solves the ground state of the Schrödinger equation without further approximation, offering an efficient solution for quantum effect modeling. Additionally, a compact mobility model considering ballistic transport is developed to capture the ballistic length dependence of mobility and the velocity overshoot effect in short-channel devices. Finally, a study on a nanosheet FET using SEMIDV is conducted. I analyze the electrical characteristics of a state-of-the-art GAA/RibbonFET with a 6 nm gate length and discuss the effects of velocity overshoot and quantum confinement on currents and capacitances with new modeling insights. A design for an ultra-short-channel transistor with a gate length down to 4.5 nm with a Vdd = 0.45 V is proposed to push the boundaries of integrated circuit technology further.

## I. INTRODUCTION

TECHNOLOGY computer-aid-design (TCAD) is the key of modern semiconductor industry. Device TCAD simulates device physics, helping to understand electrostatics, electronic currents, and other quantities in a device [1-5]. A typical device simulator solves the Poisson equation and transport equation self-consistently to obtain electrostatic potential and charge density. The drift-diffusion (DD) equation is commonly chosen as the transport equation due to its simplicity and capability for large-scale device simulation [6, 7]. Other transport methods, such as Boltzmann transport [8, 9] and nonequilibrium green's function (NEGF) method [10, 11], are too time-consuming for realistic device modeling, though they account for hot electron effects and quantum transport beyond drift-diffusion.

Over the years, the semiconductor industry has continuously shrunk transistors. Shorter channel lengths reduce transistor area and increase currents, while thinner channel and oxide thickness improve gate control to support length scaling. As device dimensions reach the nanometer scale, quantum effects emerge, making the DD equation insufficient for accurate modeling. To incorporate quantum effects into the Poisson-Drift-Diffusion framework, a common approach is introducing quantum potential into the band structure. Methods such as the van Dort model [12] and the density gradient model [13-15] use macroscopic approximations without directly solving the Schrödinger equation. Beyond quantum confinement, quasi-ballistic transport is another critical factor in nanoscale devices, prompting the development of various models to integrate quasi-ballistic transport into the DD framework [16-18].

In this paper, I present SEMIDV, an open-source TCAD semiconductor device simulator implemented in Python. The simulator is based on the Poisson-Drift-Diffusion framework with novel solutions for quantum effects. Localization Landscape Theory [19] is introduced for the first time in TCAD for quantum corrections. A simple ballistic mobility model is included to extract the transmission rate, and a velocity overshoot model is developed for quasi-ballistic transport. The program provides seamless integration with the Python environment, allowing for easy scripting and data analysis. A case study on a nanosheet field-effect transistor (FET) using SEMIDV is demonstrated in this paper.

## II. DEVICE SIMULATION FRAMEWORK

SEMIDV is a finite difference Poisson-Drift-Diffusion solver. It uses Gummel iteration [20] to self-consistently solve the coupled physics equations, with the solving flow shown in Fig. 1.

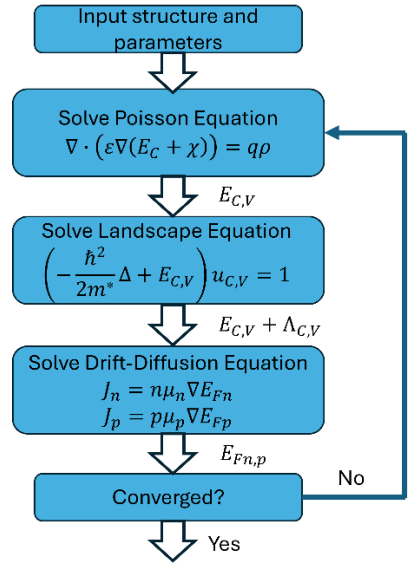

Fig. 1. The flow chart of solving physics equations in SEMIDV.

### A. Poisson Equation

Poisson equation governs the band bending in the presence of charges. (1) gives the basic form of Poisson equation where $\varepsilon$ is the permittivity, $\varphi$ is the electrostatic potential, and $\rho$ is the charge density.

$$\nabla \cdot (\varepsilon \nabla \varphi) = -\rho \quad (1)$$

By using the fact that $\varphi = -(E_C + \chi)/q$, (1) can be rewritten as (2) where $E_C$ is the conduction band energy, $\chi$ is the electron affinity, $q$ is the elementary charge, n is the electron density, p is the hole density, $N_A^-$ is the ionized acceptor, and $N_D^+$ is the ionized donor. n and p can be expressed as (3) by assuming parabolic energy band where $F_{1/2}$ is the Fermi-Dirac integral of an order of ½, $N_C$ and $N_V$ are the effective density of state for conduction and valence band respectively, $E_{Fn}$ and $E_{Fp}$ are the electron and hole fermi level respectively, $k_B$ is the Boltzmann constant, and $T$ is the temperature.

$$\nabla \cdot (\varepsilon \nabla (E_C + \chi)) = -q^2 (n - p + N_A^- - N_D^+) \quad (2)$$

$$n = N_C F_{1/2}\left(\frac{E_{Fn} - E_C}{k_B T}\right) \quad (3a)$$

$$p = N_V F_{1/2}\left(\frac{E_V - E_{Fp}}{k_B T}\right) \quad (3b)$$

The equation is nonlinear. Therefore, to solve $E_C$ with given $E_{Fn}$ and $E_{Fp}$, we need to linearize it with Taylor series and solve for the error. The linearized Poisson equation is shown as (4) where $\Delta E_C$ is the difference between previous $E_C$ and the new calculated $E_C$. All other quantities with a subscript old are calculated with previous $E_C$ value. $dn$ and $dp$ are the derivatives of $n$ and $p$ as (5) shown with a Fermi-Dirac integral of an order of -½. The simulation will converge till the error in (6) smaller than the tolerance.

$$\nabla \cdot (\varepsilon \nabla (\Delta E_C + \chi)) - q^2 dn_{old} \Delta E_C - q^2 dp_{old} \Delta E_C$$
$$= -\nabla \cdot (\varepsilon \nabla (E_{C\_old} + \chi)) - q^2 (n_{old} - p_{old} + N_A^- - N_D^+) \quad (4)$$

$$dn = \frac{N_C}{k_B T} F_{-1/2}\left(\frac{E_{Fn} - E_C}{k_B T}\right) \quad (5a)$$

$$dp = \frac{N_V}{k_B T} F_{-1/2}\left(\frac{E_V - E_{Fp}}{k_B T}\right) \quad (5b)$$

$$error = \max(|\Delta E_C|/q) \quad (6)$$

*B. Continuity Equation*

Continuity equation determines the current flow in a device. (7a) and (7b) describe the $n$ and $p$ change by the current and generation (G), and recombination (R). In SEMIDV, the transport model uses DD equations. When the Einstein relation holds, DD equation can be written as (8) where $\mu_n$ and $\mu_p$ are the electron and hole mobilities. To solve (7), I choose to solve $E_{Fn}$ and $E_{Fp}$ with Slotboom variables [7] instead of $n$ and $p$ using the Scharfetter-Gummel method [6] to not worry about heterojunction [3]. By defining Slotboom variables $\phi_n$ and $\phi_p$ as (9), (8) can be rewritten as (10). Solving for $\phi_n$ and $\phi_p$ can ensure the stability of the solution. It is important to notice that (10) is still a nonlinear function of $E_{Fn}$ and $E_{Fp}$ due to $\gamma_n$ and $\gamma_p$ in (11). However, during the Gummel iteration, I treat $\gamma_n$ and $\gamma_p$ as known values from previous iteration and solve it until the simulation converges.

$$\frac{dn}{dt} = \frac{\nabla \cdot J_n}{q} + G - R \quad (7a)$$

$$\frac{dp}{dt} = -\frac{\nabla \cdot J_p}{q} + G - R \quad (7b)$$

$$J_n = p\mu_n \nabla E_{Fn} \quad (8a)$$

$$J_p = p\mu_p \nabla E_{Fp} \quad (8b)$$

$$\phi_n = \exp\left(\frac{E_{Fn}}{k_B T}\right), \; \phi_p = \exp\left(\frac{-E_{Fp}}{k_B T}\right) \quad (9)$$

$$J_n = k_B T \mu_n N_C \gamma_n \nabla \phi_n \quad (10a)$$

$$J_p = -k_B T \mu_p N_V \gamma_p \nabla \phi_p \quad (10b)$$

$$\gamma_n = F_{1/2}\left(\frac{E_{Fn} - E_C}{k_B T}\right) / \exp\left(\frac{E_{Fn}}{k_B T}\right) \quad (11a)$$

$$\gamma_p = F_{1/2}\left(\frac{E_V - E_{Fp}}{k_B T}\right) / \exp\left(\frac{-E_{Fp}}{k_B T}\right) \quad (11b)$$

*C. Mobility Model*

Drift-diffusion model is a near-equilibrium approximation of the Boltzmann transport equation. To consider field-dependent scattering and high-field effect, mobility is usually treated as an empirical function. Commercial TCADs use complex mobility models with many parameters [21-23]. However, this leads to difficulties in model calibrations. Thus, I use a compact model approach (12) to model DD mobility $\mu_{dd}$ [24] where $\mu_0$ is the low-field mobility, UA and EU are the phonon scattering parameters, and UD, UCS, and $n_{ref}$ are Coulomb scattering parameters. To account for quasi-ballistic transport, an empirical ballistic mobility is used [25] as (13) and (14) where $\mu_b$ is the ballistic mobility. $\mu_b$ depends on channel length L which is a model parameter. $v_T$ is the thermal velocity.

$$\mu_{dd} = \frac{\mu_0}{1 + UA \cdot E_\perp^{EU} + 2 \times UD / \left(1 + n/n_{ref}\right)^{UCS}} \quad (12)$$

$$\mu_{eff}^{-1} = \mu_b^{-1} + \mu_{dd}^{-1} \quad (13)$$

$$\mu_b = \frac{v_T L}{2 k_B T / q} \quad (14)$$

For the high-field saturation model, the Caughey-Thomas model [26] is applied as (15) for $n$. $v_{sat}$ is the saturation velocity and $\beta$ is a fitting parameter. For $p$, it is analogous.

$$\mu_n = \frac{\mu_{eff}}{\left[1+\left(\frac{\mu_{eff}|\nabla E_{Fn}|}{v_{sat}}\right)^\beta\right]^{1/\beta}} \quad (15)$$

In addition, Monte Carlo simulations [27, 28] have shown that a larger $v_{sat}$ than its long-channel value is needed to model velocity overshoot. This can be understood since, without scattering, the drain-end velocity can exceed the long-channel value. Granzner et al. [27] provided a simple empirical model to fit the $v_{sat}$ for different channel lengths. Here, I will rederive that equation and provide a physical interpretation.

Velocity saturation happens when carriers encounter optical phonon scattering. The scattering rate can be modeled by the reflection probability $r$ which is expressed as (16a) where $\lambda_{opt}$ is the scattering length for optical phonons and $l$ is the length of the high-field region which is a fraction of the total length $\alpha L$. $v_{sat}$ is assumed to be inversely proportional to $r$. When $r$ is 1, optical phonon scattering is certain, and the velocity reaches its long-channel value $v_{sat0}$. As the channel length decreases, the probability of scattering decreases, causing the average drain-end velocity to increase, which is modeled by (16b). The rate of velocity increase is captured by a fitting parameter $\lambda_{sat}$.

$$r = \frac{l}{\lambda_{opt}+l} = \frac{\alpha L}{\lambda_{opt}+\alpha L} \quad (16a)$$

$$v_{sat} = \frac{v_{sat0}}{r} = v_{sat0}\left(1+\frac{\lambda_{opt}/\alpha}{L}\right) = v_{sat0}\left(1+\frac{\lambda_{sat}}{L}\right) \quad (16b)$$

### D. Localization Landscape

Quantum effects have become important in nanoscale devices. The most rigorous way is to solve the Schrödinger equation (17). However, solving (17) either using eigenvalue solver or NEGF [10] is highly time-consuming. On the other hand, methods like density gradient theory [15] provide a faster approximation for quantum potential but it still requires fitting to match quantum results.

Filoche et al. developed a novel approach to calculating quantum potential using localization landscape (LL) theory [19]. LL directly solves for the ground state energy in (17) using (18), yielding highly accurate results compared to quantum solvers without requiring special treatment [19]. (18) can be efficiently solved using a linear solver where $\Lambda_C$ is the quantum potential of electrons. This quantum potential is then used to evaluate $n$ with quantum corrections (19). The quantum potential for holes follows a similar approach.

$$\left(-\frac{\hbar^2}{2m^*}\Delta+V\right)\psi = E\psi \quad (17)$$

$$\left(-\frac{\hbar^2}{2m^*}\Delta+E_C\right)u_C = 1 \quad (18a)$$

$$W_C = \frac{1}{u_C},\ \Lambda_C = W_C - E_C \quad (18b)$$

$$n = N_C F_{1/2}\left(\frac{E_{Fn}-E_C-\Lambda_C}{k_B T}\right) \quad (19)$$

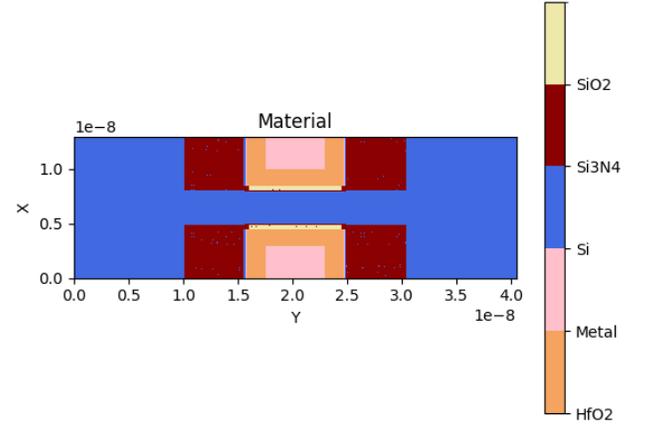

**Fig. 2.** The nanosheet FET structure calibrated to [29].

### III. A Case Study on Nanosheet FET

A case study on a nanosheet FET is performed to demonstrate the ability of SEMIDV. SEMIDV uses a Python script to construct the device structure, and its built-in visualizer helps analyze the device's properties. All figures are generated using the SEMIDV visualizer and Python. Fig. 2 the device structure designed to calibrate published nanosheet data [29], with key parameters listed in Table I.

TABLE I
PARAMETERS OF NANOSHEET

| Parameter | Value |
|---|---|
| Gate Length | 6 nm |
| Channel Thickness | 3 nm |
| Spacer Length | 6 nm |
| SiO$_2$ Thickness | 0.5 nm |
| High-K Thickness | 1.5 nm |
| Channel Doping | $1\times10^{15}\ cm^{-3}$ |
| S/D Doping | $2\times10^{20}\ cm^{-3}$ |

### A. Transport Characteristics

Fig. 3 shows the calibrated I-V curves compared to [29]. Based on the reported apparent mobility in [29], $\mu_0$ is chosen to be $310\ cm^2V^{-1}s^{-1}$. For ballistic mobility, the electrical channel length is estimated to be 9 nm from the band diagram in Fig. 4 with $\mu_b = 208\ cm^2V^{-1}s^{-1}$, suggesting approximate 60% of transmission. Now, let's examine $v_{sat}$. $v_{sat} = 3.7\times10^7\ cm/s$ is chosen to model the high-drain-bias current larger than the typical long-channel $v_{sat0} = 2\times10^7\ cm/s$ reported in [27]. This indicates that at the drain end, electrons travel with very high ballisticity. In Fig. 4, the drain-high-field region is about 5 nm, allowing us to estimate $\lambda_{opt}$ to be about 4.25 nm, resulting in a transmission rate of about 46% in the high-field region. In our simulation (Fig. 5), the peak velocity is $3.57\times10^7\ cm/s$ and the injection velocity is $1.38\times10^6\,cm/s$ which aligns with the values in [29].

Next, I will discuss the effect of velocity on capacitance. A comparison of intrinsic gate capacitances (Cggi) between $v_{sat}=3.7\times10^7\,cm/s$ and $2\times10^7\,cm/s$ is shown in Fig. 6. It is obvious that the high-drain-bias (V$_{dsat}$) capacitance of $2\times10^7\,cm/s$ is larger than $3.7\times10^7\,cm/s$. Due to the current continuity, the case with larger $v_{sat}$ will have fewer electrons in the drain end resulting in lower capacitance than typical velocity saturation limits. In more aggressive case (Fig. 9), the V$_{dsat}$ Cggi will be less than 2/3 of the V$_{dlin}$ Cggi which is the theorical limits for long-channel FETs. This finding suggests the need for modifications to transistor compact models in quasi-ballistic regimes to ensure accurate IC designs.

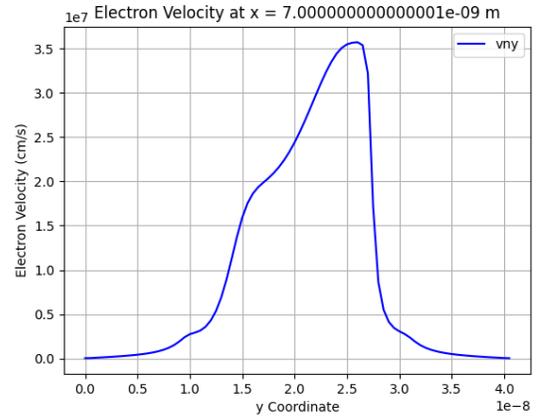

**Fig. 5.** Simulated velocity profile from SEMIDV.

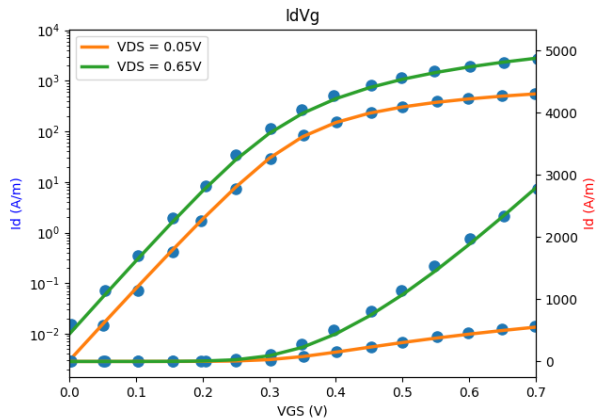

**Fig. 3.** The SEMIDV calibrated IdVg curves versus the Intel 6nm RibbonFET [29]. The lines are from SEMIDV and the symbols are from [29].

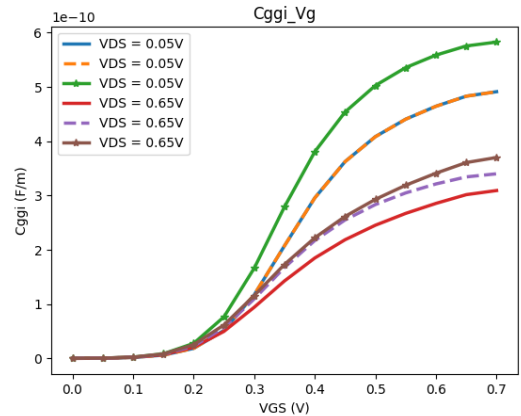

**Fig. 6.** Simulated intrinsic gate capacitance (Cggi). Solid lines are the baseline. Dash lines are for $v_{sat}=2\times10^7\,cm/s$. Symbolled lines are the case without quantum effect.

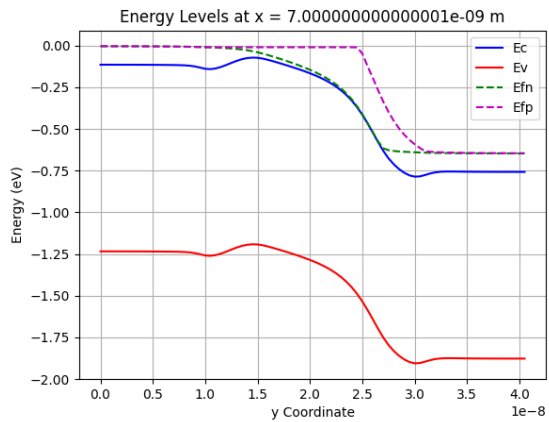

**Fig. 4.** Simulated band-diagram from SEMIDV.

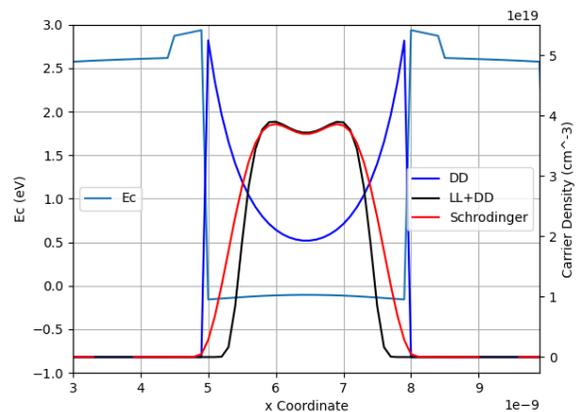

**Fig. 7.** The simulated electron distribution in the channel by DD, LL+DD, and the Schrödinger equation.

*B. Quantum Confinement*

I use LL theory to directly compute the ground state from the Schrödinger equation. Fig. 7 compares the electron concentration with and without the quantum correction, as well as the solution from the Schrödinger equation. Without quantum correction, electrons accumulate on the surface. With LL, electrons are confined to the middle of the channel, similar to the Schrödinger equation. The effects of this quantum confinement are reflected in gate capacitance, as shown in Fig. 6. Cggi becomes smaller due to quantum confinement, which can be understood by considering an equivalent quantum capacitance in series with the oxide capacitance.

In addition to the vertical confinement, LL theory can also model the quantum effect in the lateral direction. From Fig. 4, $E_C$ lowering in source and drain is commonly seen in quantum simulations [11, 30] due to the carrier reflection at the source/drain barrier.

*C. Design of an ultra-short-channel transistor*

With the calibrated SEMIDV model, I would like to propose a design for the ultra-short-channel transistor. To improve the gate control and reduce short-channel effects, we need to reduce channel thickness and increase gate capacitance. However, aggressively reducing thickness will cause mobility degradation due to surface roughness scattering. Based on the data in [29], I choose a 2.5 nm channel thickness where the extracted DD mobility is $282\ cm^2V^{-1}s^{-1}$.

To boost gate capacitance, it is not possible to do geometry enhancement to the thickness in TABLE I. Fortunately, Park et al. demonstrated using ferroelectric material to reduce equivalent oxide thickness (EOT) [31]. With ferroelectric superlattice, they can reduce the EOT from 8 Å for the interfacial $SiO_2$ to 6.5 Å, which shows an equivalent -1.5 Å for the ferroelectric layer. In SEMIDV, there is no model for such "negative capacitance". Since the negative capacitance effect only influences the oxide layer not the Silicon, we can simply adjust the dielectric constant to mimic the behavior. Based on the -1.5 Å reported in [31], I choose the EOT to be 5 Å - 1.5 Å = 3.5 Å.

With these enhancements, I design a transistor with 4.5 nm gate length and 2.5 nm channel thickness with all the other parameters same as TABLE I. Fig. 8 compares the new transistor with the Intel RibbonFET [29]. The designed FET has a better subthreshold swing (SS) and drain-induced-barrier-lowering (DIBL) even at shorter length due to reduced thickness and EOT. The linear-region current is lower because of the degraded mobility. However, the saturation current is increased for two reasons. First, the gate capacitance boost (Fig. 9) increases the charges in the channel. Second, the reduced gate length improves the ballisticity leading to a higher injection velocity. It is also interesting to notice that the $V_{dsat}$ Cggi drops to 1/2 of the $V_{dlin}$ Cggi which is lower than the long-channel limits (2/3) due to high drain velocity.

Finally, to test how low the supply voltage can be, I match the on and off current to IRDS 2023 [32]. It is found that the designed FET can reduce Vdd down to 0.45 V (Fig. 8). The intrinsic delay (CV/I) is estimated to be 0.25 ps with switching energy per width ($CV^2$) being 0.9 pJ/m.

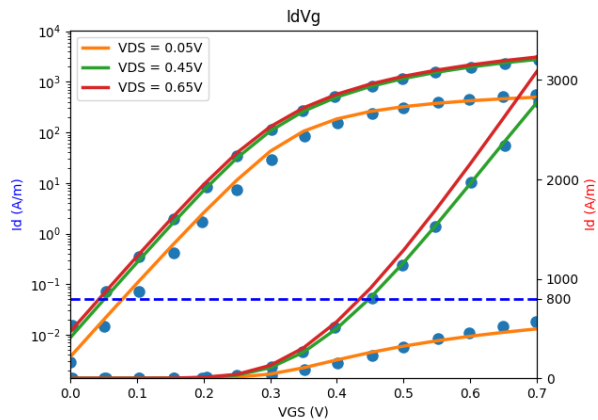

**Fig. 8.** IdVg curves of the proposed FET at 0.05, 0.45, and 0.65 $V_{DS}$. The symbols are the data from [29] at 0.05 and 0.65 $V_{DS}$. The on-current level is set to be 800 A/m.

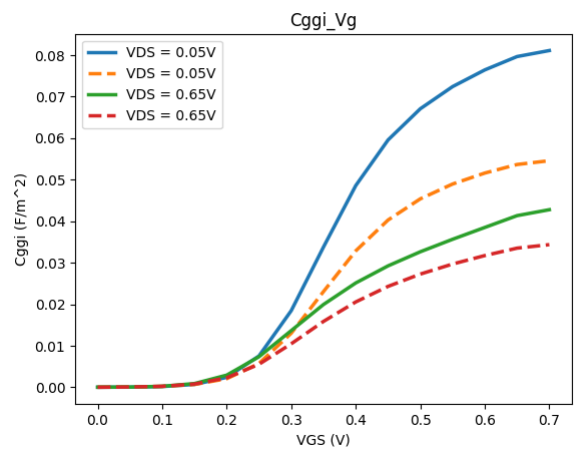

**Fig. 9.** The intrinsic gate capacitance (Cggi) for the designed 4.5 nm FET (solid lines) versus the calibrated 6 nm FET (dash lines). A significant capacitance boost comes from the low EOT of 4.5nm FET.

## V. Conclusion

SEMIDV is a simple and compact device simulator that allows us to study nanoscale devices operating in the quantum regime. With SEMIDV, I examine the characteristics of ultra-short-channel FETs and provide insights into both modeling and design for devices in quantum limits. With channel length, thickness, and EOT scaling, we can push the transistor to a few nanometers long with Vdd as low as 0.45 V.